\documentclass[aps,prd,groupedaddress,preprint,eqsecnum,nofootinbib]{revtex4}
\usepackage{amsmath}
\usepackage{graphicx,epsf}
\usepackage[usenames]{color}

\newif\ifdraft
\drafttrue
\newif\ifpreprint
\preprinttrue

\def\spa#1.#2{\left\langle#1\,#2\right\rangle}
\def\spb#1.#2{\left[#1\,#2\right]}
\def\spash#1.#2{\spa{\smash{#1}}.{\smash{#2}}}
\def\spbsh#1.#2{\spb{\smash{#1}}.{\smash{#2}}}
\def\sand#1.#2.#3{%
\left\langle\smash{#1}{\vphantom1}^{-}\right|{#2}%
\left|\smash{#3}{\vphantom1}^{-}\right\rangle}
\def\sandpp#1.#2.#3{%
\left\langle\smash{#1}{\vphantom1}^{+}\right|{#2}%
\left|\smash{#3}{\vphantom1}^{+}\right\rangle}
\def\sandpm#1.#2.#3{%
\left\langle\smash{#1}{\vphantom1}^{+}\right|{#2}%
\left|\smash{#3}{\vphantom1}^{-}\right\rangle}
\def\sandmp#1.#2.#3{%
\left\langle\smash{#1}{\vphantom1}^{-}\right|{#2}%
\left|\smash{#3}{\vphantom1}^{+}\right\rangle}

\def\Neqfour{{{\cal N}=4}}

\def\Neqeight{{{\cal N}=8}}
\def\NeqEight{{{\cal N}=8}}

\def\be{\begin{equation}}
\def\ee{\end{equation}}
\def\bea{\begin{eqnarray}}
\def\eea{\end{eqnarray}}
\def\ba{\begin{eqnarray}}
\def\ea{\end{eqnarray}}

\def\M{{\cal M}}
\def\N{{\cal N}}

\def\tlambda{{\tilde\lambda}}

\newbox\charbox
\newbox\slabox
\def\s#1{{      
        \setbox\charbox=\hbox{$#1$}
        \setbox\slabox=\hbox{$/$}
        \dimen\charbox=\ht\slabox
        \advance\dimen\charbox by -\dp\slabox
        \advance\dimen\charbox by -\ht\charbox
        \advance\dimen\charbox by \dp\charbox
        \divide\dimen\charbox by 2
        \raise-\dimen\charbox\hbox to \wd\charbox{\hss/\hss}
        \llap{$#1$} }}

\DeclareMathAlphabet{\mathpzc}{OT1}{pzc}{m}{it}

\begin{document}

\title{A Note on Single Soft Scalar Emission of $\NeqEight$ SUGRA
and $E_{7(7)}$ Symmetry}

\author{Song He}
\email{songhe@aei.mpg.de}
\author{Hua Xing Zhu}
\email{hxzhu@pku.edu.cn}

\affiliation{School of Physics, Peking University, Beijing, 100871,
China }

\bigskip

\begin{abstract}

We study single soft scalar emission amplitudes of $\Neqeight$
supergravity (SUGRA) at the one-loop level using an explicit formula
for one-loop amplitudes in terms of tree amplitudes, which in turn
are evaluated using supersymmetric BCFW recursion relations. It
turns out that the infrared-subtracted amplitudes vanish in the soft
momentum limit, which supports the conjecture that $E_{7(7)}$
symmetry has no anomalies at the one-loop level.

\end{abstract}


\maketitle

\newpage
\section{Introduction}
There has been a renewed interest in $\Neqeight$ supergravity
(SUGRA) in recent years. In~\cite{conjecture}, the authors made the
bold conjecture that the theory may be ultraviolet(UV) finite up to
all orders in perturbation theory. To support the conjecture, they
provided all-loop evidence for the finiteness by promoting the
``no-triangle hypothesis''\footnote{Note that for one-loop
$\Neqeight$ SUGRA amplitudes, the absence of triangle, bubble and
rational terms has been proven, and the use of this terminology is
merely a convention.}~\cite{notriangle} to higher loops and using
string duality arguments in \cite{Green:2006gt}. Since then, various
higher loop calculations have confirmed explicitly that $\Neqeight$
SUGRA in four dimensional spacetime is UV finite at three
loops~\cite{finiteness3}, and very recently at four
loops~\cite{finiteness4}. There are also string theory arguments in
favor of the finiteness of $\Neqeight$ SUGRA
~\cite{Green:2006gt,string}.

On the other hand, it has long been known that on-shell classical
$\Neqeight$ SUGRA has a local $SU(8)$ symmetry and a hidden global
$E_{7(7)}$ symmetry~\cite{CJ,WN}. Before gauge-fixing, $E_{7(7)}$ is
linearly realized and acts on 133 scalars as well as vectors present
in the classical action, independent from the local $SU(8)$
symmetry. The 63 local parameters of $SU(8)$ can be made used of to
remove 63 non-physical scalars, leaving 70 massless scalars, which
leads to an non-linearly realization of $E_{7(7)}$ on the remaining
scalars. The action of the non-linearly realized $E_{7(7)}$ on
$\Neqeight$ SUGRA fields was only revealed recently~\cite{explicit},
exact to all orders in gravitational coupling constant. It is
possible but still not clear that this hidden $E_{7(7)}$ is relevant
to the conjectured finiteness of $\Neqeight$ SUGRA.

Recently, the emission of a single soft scalar in $\Neqeight$ SUGRA
tree amplitudes was examined in~\cite{BEF}, in order to find the
imprint of $E_{7(7)}$ symmetry, as expected from low energy theorem
associated with soft Goldstone boson emission. The amplitudes for a
single soft scalar emission were found to vanish generally, and it
should be noted that the result is beyond the expectation from low
energy theorem in pion physics, where a single soft pion emission is
generally non-vanishing and can be obtained from the sum of Feynman
diagrams in which the soft pion is attached to other external
lines~\cite{Bando:1987br}. This is due to the fact that in the
diagrams where the soft Goldstone boson is attached to external
particles, taking the soft limit could lead to propagator
singularities when the external particles are
on-shell~\cite{Weinberg:1996kr}. In $\Neqeight$ SUGRA case there are
cubic vertices in the Lagrangian through which a soft scalar could
attach to external particles. However, these vertices vanish by
themselves and overcompensate the propagator singularities. The same
result has been obtained in~\cite{ACK}. By generalizing the BCFW
recursion relations~\cite{BCFW} to $\Neqfour$ supersymmetric
Yang-Mills theory and $\Neqeight$ SUGRA, the authors of~\cite{ACK}
related general tree amplitudes to three-particle amplitudes, which
vanish fast enough to overcompensate the propagator singularities in
the soft limit, thus established that amplitudes with single soft
scalar emission vanish generally. They also found that double soft
scalar emission amplitudes can be related to commutator of two
``broken generators'' which label the soft scalars. In~\cite{KK},
the footprint of $E_{7(7)}$ symmetry on tree amplitudes was examined
from a very different perspective. Considering the consequences of
Noether current conservation associated with $E_{7(7)}$ symmetry, it
turns out that single soft scalar emission amplitudes can be related
to amplitudes without soft scalar, but with extra ``axial'' charge
attached to external particles. The result shows that such ``axial''
charge vanishes at the tree level. As was explained in~\cite{KK},
when combining with the results of~\cite{BEF,ACK}, this establishes
the result of low energy theorem for an $E_{7(7)}$ symmetry at the
tree level.

Given the low energy theorem of $E_{7(7)}$ symmetry at the tree
level, it is very natural and interesting to see if this symmetry
persists at the higher-order level. From the fact that chiral
$SU(8)$ one-loop triangle anomalies vanish~\cite{SU8anomaly}, it is
expected that $E_{7(7)}$ is not anomalous at least at the one-loop
level. The authors of~\cite{KLR} have established the low energy
theorem for one-loop $n$-point amplitudes, by assuming the
$E_{7(7)}$ symmetry at the one-loop level. They found that the soft
limit of the bosonic 4-point amplitudes vanish for complex momenta
and the ``axial'' charge vanishes for all one-loop amplitudes. It
remains to examine the soft limit of one loop $n$-point ($n\geq 5$)
amplitudes, in order to confirm the low energy theorem of $E_{7(7)}$
symmetry at the one-loop level.

In this note we make a first attempt towards this goal. Our
perspective is different from that of~\cite{KLR} but closer to that
of~\cite{ACK}. We shall use a simple formula for general one-loop
amplitude in terms of tree amplitudes~\cite{ACK} follows from ``no
triangle hypothesis''~\cite{notriangle}. In section~\ref{susybcfw},
we review supersymmetric BCFW recursion relations and the vanishing
result of single soft scalar emission at the tree level. Then the
result is generalized to the one-loop level for the
infrared-subtracted amplitudes in section~\ref{onelooplevel}, which
can be viewed as a strong evidence for the absence of $E_{7(7)}$
anomalies. Conclusion and Discussions are presented in the end.

\section{Single Soft Scalar Emission of $\Neqeight$ SUGRA
at the Tree Level} \label{susybcfw}

It has been argued from different perspectives~\cite{BEF,ACK,KK,KLR}
that tree amplitudes with a single soft scalar emission in
$\Neqeight$ SUGRA vanish, which can be viewed as hints of a hidden
$E_{7(7)}$ symmetry. Besides, double soft scalar emission has been
calculated in~\cite{ACK} to reveal the non-trivial structure of
$E_{7(7)}$. The key point leads to this result in~\cite{ACK} is the
generalization of BCFW recursion relations~\cite{BCFW} to maximally
supersymmetric theories which we review below, and we refer
to~\cite{ACK,Brandhuber:2008pf} for details.

One beautiful insight in
\cite{AK,ACK} is that amplitudes of particles with higher spin have
better large $z$ scaling, where $z$ is associated with the BCFW
deformation
\begin{equation}
\label{BCFWdeform}
 \lambda_1(z)=\lambda_1+z\lambda_2,\qquad
\tlambda_2(z)=\tlambda_2-z\tlambda_1.
\end{equation}
In addition, maximal SUSY can relate all the helicity states in a
CPT invariant super-multiplet to each other, which allows one to
label the external states in a natural, continuous way. It's
realized as follows. The external states are represented by
Grassmann coherent states $|\eta\rangle$ or $|\bar{\eta}\rangle$,
which diagonalize not only the momentum but also the supercharge
$Q_I$ or $\bar{Q}^I$, respectively. For a massless particle with
momentum $(\sigma^\mu p_\mu)_{\alpha\dot{\alpha}}= \lambda_\alpha
\tlambda_{\dot{\alpha}}$, the Grassmann coherent states are defined
as
\begin{equation}
\label{coherentstates}
 |\bar{\eta}, \lambda, \tlambda \rangle = e^{\bar{Q}^{I
\dot{\alpha}}\tilde{w}_{\dot{\alpha}} \bar{\eta}_I} |+s ,\lambda,
\tlambda \rangle, \quad |\eta,\lambda, \tlambda\rangle = e^{Q_{I
\alpha} w^{\alpha} \eta^I}|-s,\lambda, \tlambda\rangle
\end{equation}
where $w_\alpha$ and $\tilde w_{\dot a}$ are spinors such that
$\langle w,\lambda\rangle = 1$ and $[\tilde{w},\tlambda] = 1$. Note
that $w_\alpha$ and $\tilde w_{\dot a}$ are not uniquely defined,
but up to an additive shift, e.g. $w_\alpha\sim w_\alpha
+c\lambda_\alpha$. The Grassmann coherent states defined in
Eq.~(\ref{coherentstates}) are built from the highest spin states
$|+s,\lambda,\tlambda\rangle$ and $|-s,\lambda,\tlambda\rangle$,
respectively, where $Q|+s\rangle=\bar{Q}|-s\rangle=0$.

Note that $\eta$ and $\bar{\eta}$ are equally valid descriptions of
the complete supermultiplet; they are related to each other by a
Grassmann Fourier transformation
\begin{equation}
|\bar{\eta}\rangle=\int~d^\N\eta e^{\eta\bar{\eta}}|\eta\rangle,
\qquad |\eta\rangle=\int~d^\N\bar{\eta}
e^{\bar{\eta}\eta}|\bar{\eta}\rangle.
\end{equation}
Now all the amplitudes can be expressed as smooth functions of
$\eta$ and $\bar{\eta}$
\begin{equation}
 \M(\{ \eta_i,\lambda_i,\tlambda_i\};\{ \bar{\eta}_{\bar{i}},
\lambda_{\bar{i}},\tlambda_{\bar{i}}\})
\end{equation}
In terms of amplitudes in $\eta$ representation, the BCFW recursion
relations are generalized to~\cite{ACK,Brandhuber:2008pf}
\begin{eqnarray} && \M(\{\eta_1(z),\lambda_1(z),\bar
\lambda_1\},\{\eta_2,\lambda_2,\bar \lambda_2(z)\}, \eta_i) =
\nonumber
\\
 && \sum_{L,R}\int d^{\cal N} \eta
\M_L(\{\eta_1(z_P),\lambda_1(z_P),\bar\lambda_1\}, \eta, \eta_L) \,
\frac{1}{P^2(z)} \,
\M_R(\{\eta_2,\lambda_2,\bar\lambda_2(z_P)\},\eta,\eta_R)
\label{fullsusy}
\end{eqnarray}
where $\eta_1(z_P)=\eta_1+z_P \eta_2$ is the supersymmetric
counterpart of BCFW deformation Eq.~(\ref{BCFWdeform}).

Now the vanishing result of single soft scalar emission in
$\Neqeight$ SUGRA can be derived from supersymmetric BCFW recursion
relations. Starting with any amplitude containing a particle with
momentum $p$ in $\eta$ or $\bar{\eta}$ representation, denoted as
$\M(\eta,...)$ or $\M(\bar{\eta},...)$, the single soft scalar
emission is obtained by
\begin{equation}
\lim_{p\rightarrow 0}\int d^8\eta \eta^{abcd}\M(\eta,...),\
\textrm{or}\ \lim_{p\rightarrow 0}\int d^8\bar{\eta}
\bar{\eta}_{abcd}\M(\bar{\eta},...),
\end{equation}
where we first multiply the amplitude by $\eta^{abcd}\equiv \eta^a
\eta^b \eta^c \eta^d$ or $\bar{\eta}_{abcd}\equiv \bar{\eta}_a
\bar{\eta}_b \bar{\eta}_c \bar{\eta}_d$ with (sub)superscripts in
(anti-)fundamental representations of $SU(8)$, and integrate it over
$\eta$ or $\bar{\eta}$, forcing the corresponding particle to be a
scalar, then take the soft limit $p\rightarrow 0$. In terms of
spinors, the soft limit is not uniquely defined. If we want to use
$p\sim \delta $ and then take $\delta\rightarrow 0$, we can take
$\lambda\sim \delta^\alpha$ and $\tilde{\lambda}\sim
\delta^{1-\alpha}$ for any $\alpha\in [0,1]$, and \textbf{the} soft
limit should be given by the largest contribution with a certain
$\alpha$. In the following we use $f(p)\sim \mathcal{O}(g(\delta))$
to express that when $p\sim \delta\rightarrow 0$, $f(p)$ is of the
same order as or higher order than $g(p)$ in $\delta$, i.e.
\begin{equation}
\lim_{p\rightarrow 0} \frac{f(p)}{g(p)}=c,
\end{equation}
where it is understood as \textbf{the} soft limit if $f(p)$ is
expressed as a function of $\lambda$ and $\tilde{\lambda}$. Besides,
$c$ is a finite constant which can be zero.

To begin our discussion, we recall that the three particle amplitude
can be either holomorphic or anti-holomorphic,
\begin{equation}\label{M3}
\M_3(\eta_1,\eta_2,\eta_3)=\frac{\delta^{16}(\sum^{3}_{i=1}\tilde{\lambda}_{i}\eta_{i})}{(\spb
1.2 \spb 2.3 \spb 3.1)^2},\ \textrm{or}\ \prod^{3}_{i=1}\int
d^8\bar{\eta}_i
\exp(\eta_{i}\bar{\eta}_{i})\frac{\delta^{16}(\sum^{3}_{i=1}\lambda_{i}\bar{\eta_{i}})}{(\spa
1.2 \spa 2.3 \spa 3.1)^2},
\end{equation}

It is instructive to take a close look on the effect of taking soft
limit in different ways. For a general three particle amplitude
$\M_3(1,2,3)$, we take the momentum of the first particle to be soft
$p_1\sim\delta$ by taking $\lambda_1\sim\delta^\alpha$ and
$\tlambda_1\sim\delta^{1-\alpha}$ for any $\alpha\in [0,1]$. The
anti-holomorphic part of $\M_3$ is
\begin{equation}
 \M^{ah}_3=\frac{\delta^{16}(\tlambda_i\eta_i)}
{(\spb 1.2 \spb 2.3 \spb 3.1)^2}
\end{equation}
By momentum conservation
$\tlambda_2=-\delta^\alpha\tlambda_1-\tlambda_3$,
we have
\begin{equation}
\label{antiholo}
 \M^{ah}_3=\delta^{2-4\alpha}{\spb \hat{1}.3}^2
\delta^8(\eta_1-\delta^\alpha\eta_2)\delta^8(\eta_3-\eta_2),
\end{equation}
where we have rescaled the spinor of particle 1 as
$\lambda_1=\delta^\alpha\hat{\lambda}_1$ and
$\tlambda_1=\delta^{1-\alpha}\hat{\tlambda}_1$, where
$\hat{\lambda}_1$ and $\hat{\tlambda}_1$ are hard spinors. The
holomorphic part of $\M_3$ is
\begin{eqnarray}
 \M^h_3&=&\prod_i\int\,d^8\bar{\eta}_i e^{\bar{\eta}_i\eta_i}
\frac{\delta^{16}(\lambda_i\bar{\eta}_i)} {(\spa 1.2 \spa 2.3 \spa
3.1)^2}\nonumber
\\
\label{holo} &=&\prod_i\int\,d^8\bar{\eta}_i e^{\bar{\eta}_i\eta_i}
\delta^{-2+4\alpha}{\spa \hat{1}.3}^2
\delta^8(\bar{\eta}_1-\delta^{1-\alpha}\bar{\eta}_2)
\delta^8(\bar{\eta}_3-\bar{\eta}_2)
\end{eqnarray}

Therefore, for a soft particle other than scalar, 
the soft limit of three particle amplitude depends on $\alpha$, i.e.
it depends on the way to take soft limit. However, for soft scalar
limit which needs the delta function to provide exactly four
components of $\eta_1$ or $\bar{\eta}_1$, the delta function
contributes $\delta^{4\alpha}$ for anti-holomorphic part, and
$\delta^{4-4\alpha}$ for holomorphic part, which yields a total
contribution of the order $\delta^2$ for whatever $\alpha$ and
either holomorphic or anti-holomorphic case,
\begin{equation}\label{singletree3}
\int d^8\eta_1 \eta_1^{abcd}\M_3(\eta_1,\eta_2,\eta_3)\sim
\mathcal{O}(\delta^2)\rightarrow 0 \ \textrm{as}\  p_1\sim \delta
\rightarrow 0.
\end{equation}

For $n+1(n\geq 3)$-point amplitude with a single soft scalar and $n$
hard particles, it is straightforward to iteratively use
supersymmetric BCFW recursion relations by deforming any two hard
particles in the same (sub-)amplitude containing the soft scalar in
every step, until there is a three particle amplitude with the
single soft scalar in the factorization, which is of the order $\sim
\mathcal{O}(\delta^2)$. However, this is accompanied by a propagator
pole since the propagator attached to this three amplitude is
\begin{equation}
\frac{1}{(p_1+p_i')^2}=\frac{1}{2p_1\cdot p_i'}\sim
\mathcal{O}(\delta^{-1}) \ \textrm{as}\  p_1\sim \delta \rightarrow
0,
\end{equation}
where $p_i'(2\leq i\leq n+1)$ is the (possibly deformed in previous
steps of decomposition) momentum of the hard external particle in
this three particle amplitude and $p_i'\sim \mathcal{O}(1)$.
Therefore, we have
\begin{equation}\label{singletree}
\int d^8\eta_1 \eta_1^{abcd}
\M_{n+1}(\eta_1,\eta_2,...,\eta_{n+1})\sim
\mathcal{O}(\delta)\rightarrow 0 \ \textrm{as}\  p_1\sim \delta
\rightarrow 0\,
\end{equation}
for $n\geq 3$.

\section{Generalization to One-loop Amplitudes}\label{onelooplevel}

The ``no-triangle hypothesis'' for one-loop amplitudes in
$\Neqeight$
SUGRA has been discovered and proved in \cite{notriangle}. It was
also proved from a different perspective in~\cite{ACK}. The absence
of triangle, bubble coefficients and rational terms leads to the
simple formula for any one-loop amplitude in $\Neqeight$ SUGRA,
which is a sum of box integrals with coefficients purely given by
products of tree amplitudes,
\begin{eqnarray}\label{1-loop}
&&\M^{1-loop}_{n}=\sum_{A,B,C,D\subset\{n\}}\sum_{l^*}\prod_{a=AB,BC,CD,DA}\int
d^8\eta_{a}\nonumber\\
&&\M_A(\eta_{DA},-l^*_{DA};A;\eta_{AB},l^*_{AB})\M_B(\eta_{AB},-l^*_{AB};B;\eta_{BC},l^*_{BC})\nonumber\\
&&\times
\M_C(\eta_{BC},-l^*_{BC};C;\eta_{CD},l^*_{CD})\M_D(\eta_{CD},-l^*_{CD};D;\eta_{DA},l^*_{DA})\nonumber\\
&&\times I_4(P_A,P_B,P_C,P_D).
\end{eqnarray}

Here the first summation is over all non-empty, non-intersecting
subsets $A,B,C$ and $D$(corners) of $n$ particles, $A\cup B\cup
C\cup D=\{n\}\equiv\{1,...,n\}$, and the second is over (generally
two) solutions of equations
\begin{equation}\label{frozen}
l^2=(l-P_B)^2=(l-P_B-P_C)^2=(l+P_A)^2=0,
\end{equation}
where $P_A=\sum_{i\in A}p_i$ and similarly for $B,C$ and $D$. The
$4\times8$ fold Grassmann integrations include those over 8
Grassmann variables for the internal line between two corners $D$
and $A$, $\eta^{I}_{DA}$ with $I=1,...,8$, and similarly for $AB,BC$
and $CD$. In addition, the corresponding momenta are denoted by
$l^*_{DA}=l^*+P_A$, $l^*_{AB}=l^*$, $l^*_{BC}=l^*-P_B$ and
$l^*_{CD}=l^*-P_B-P_C$, as presented in the four tree amplitudes
$\M_A, \M_B, \M_C$ and $\M_D$. $A$ is short for
$\{\eta_i,p_i\}$ or $\{\eta_i,\lambda_i,\tilde{\lambda}_{i}\}$ with
$i\in A$, and similarly for $B,C$ and $D$.  The product of these
four tree amplitudes is the coefficient of the box integral
$I_4(P_A,P_B,P_C,P_D)$ which is given by~\cite{Bern:1995ix}
\begin{equation}
I_4(K_1,K_2,K_3,K_4)=-\frac{r_\Gamma}{2\sqrt{\textrm{det}S}}F_4,
\end{equation}
where
$r_\Gamma=\frac{\Gamma(1+\epsilon)\Gamma^2(1-\epsilon)}{\Gamma(1-2\epsilon)}$,
the symmetric $4\times 4$ matrix $S$ is,
\begin{equation}\label{S}
S_{ij}=-\frac{1}{2}(K_i+...+K_{j-1})^2\ \textrm{for}\ i\neq j,\
S_{ii}=0,
\end{equation}
and box functions are given by,
\begin{eqnarray}\label{box}
F^{4m}(K_1,K_2,K_3,K_4) & = & \frac{1}{2}\left( -{\rm Li}_2((1-\lambda_1+\lambda_2+\rho)/2)+{\rm Li}_2((1-\lambda_1+\lambda_2-\rho)/2)\right. \nonumber \\
& & -{\rm Li}_2(-(1-\lambda_1-\lambda_2-\rho)/(2\lambda_1))+{\rm Li}_2(-(1-\lambda_1-\lambda_2+\rho)/(2\lambda_1)) \nonumber \\
& & \left. -\frac{1}{2}\ln \left(\frac{\lambda_1}{\lambda_2^2}\right)\ln\left( \frac{1+\lambda_1-\lambda_2+\rho}{1+\lambda_1-\lambda_2-\rho} \right) \right) , \nonumber \\
F^{3m}(k_1,K_2,K_3,K_4) & = & -\frac{1}{2\epsilon^2}\left(
(-s)^{-\epsilon}+ (-t)^{-\epsilon}-(-K
_2^2)^{-\epsilon}-(-K_4^2)^{-\epsilon}\right) \nonumber \\
& & + {\rm Li}_2\left( 1-\frac{K _2^2}{s}\right)+ {\rm Li}_2\left(
1-\frac{K_4^2}{t}\right) - {\rm Li}_2\left(
1-\frac{K_2^2K_4^2}{st}\right) \nonumber \\ & & +
\frac{1}{2}\ln^2\left(\frac{s}{t}\right) -
\frac{1}{2}\ln\left(\frac{K_4^2}{s}\right)\ln\left(\frac{K_3^2}{s}\right)-
\frac{1}{2}\ln\left(\frac{K_2^2}{t}\right)\ln\left(\frac{K
_3^2}{t}\right), \nonumber \\
F^{2m\, e}(k_1,K_2,k_3,K_4) & = & -\frac{1}{\epsilon^2}\left(
(-s)^{-\epsilon}+ (-t)^{-\epsilon}-(-K_2^2)^{-\epsilon}-(-K
_4^2)^{-\epsilon}\right) \nonumber \\ & & + {\rm Li}_2\left(
1-\frac{K_2^2}{s}\right) + {\rm Li}_2\left( 1-\frac{K
_2^2}{t}\right)+ {\rm Li}_2\left( 1-\frac{K_4^2}{s}\right)+ {\rm Li}_2\left( 1-\frac{K_4^2}{t}\right)\nonumber \\
& & - {\rm Li}_2\left( 1-\frac{K_2^2K_4^2}{st}\right)+\frac{1}{2}\ln^2\left(\frac{s}{t}\right) , \nonumber \\
F^{2m\, h}(k_1,k_2,K_3,K_4) & = & -\frac{1}{2\epsilon^2}\left(
(-s)^{-\epsilon}+ 2(-t)^{-\epsilon}-(-K
_3^2)^{-\epsilon}-(-K_4^2)^{-\epsilon}\right) \nonumber \\
& & + {\rm Li}_2\left( 1-\frac{K_3^2}{t}\right)+ {\rm Li}_2\left(
1-\frac{K_4^2}{t}\right)+ \frac{1}{2}\ln^2\left(\frac{s}{t}\right)-
\frac{1}{2}\ln\left(\frac{K
_4^2}{s}\right)\ln\left(\frac{K_3^2}{s}\right), \nonumber \\
F^{1m}(k_1,k_2,k_3,K_4) & = & -\frac{1}{\epsilon^2}\left(
(-s)^{-\epsilon}+ (-t)^{-\epsilon}-(-K_4^2)^{-\epsilon}\right)
\nonumber \\ & & +{\rm Li}_2\left( 1-\frac{K_4^2}{s}\right)+{\rm
Li}_2\left(
1-\frac{K_4^2}{t}\right)+\frac{1}{2}\ln^2\left(\frac{s}{t}\right)+\frac{\pi^2}{6},
\nonumber\\
F^{0m}(k_1,k_2,k_3,k_4) & = & -\frac{1}{\epsilon^2}\left(
(-s)^{-\epsilon}+(-t)^{-\epsilon}\right)+\frac{1}{2}\ln^2\left(\frac{s}{t}\right)+\frac{\pi^2}{2}.
\end{eqnarray}
for four, three, ..., zero-mass case respectively, where $\epsilon$
is the infrared cutoff in dimensional regularization. $k_i$ are
on-shell momenta for the case when there is only one external leg in
a corner and $K_i$ off-shell momenta for more generic case. The
Mandelstam variables are $s=(k_1+k_2)^2$ and $t=(k_1+k_4)^2$ for
zero-mass case and similarly for other cases with possible off-shell
momenta. The dilogarithm function is defined as
\begin{equation}
\textrm{Li}_2(z)=-\int_{0}^{z}\frac{dx}{x}\ln(1-x),
\end{equation}
and in four-mass case $\rho$ is defined as,
\begin{equation}
\rho=\sqrt{1-2(\lambda_1+\lambda_2)+(\lambda_1-\lambda_2)^2}\
\textrm{with} \ \lambda_1=\frac{K^2_1K^2_3}{st}\ \textrm{and} \
\lambda_2=\frac{K^2_2K^2_4}{st}.
\end{equation}

Before proceeding, an important remark on infrared divergences is
needed. These box functions, except the four-mass one, all possess
infrared divergences, which are regularized by computing in
$D=4-2\epsilon$ dimension. A general one-loop amplitude then can be
expanded in powers of $\epsilon$,
\begin{equation}
\M^{1-loop}_n(\epsilon)=\frac{C_2}{\epsilon^2}+\frac{C_1}{\epsilon}+C_0+\mathcal{O}(\epsilon),
\end{equation}
where $C_2,C_1$ and $C_0$ are functions of kinematic invariants. The
result we shall present for one-loop single soft scalar emission has
two folds of meanings. First, as it stands, we shall prove that for
any one-loop amplitude with a single soft scalar,
\begin{equation}
\lim_{\delta\rightarrow 0}C_i(\delta)=0,
\end{equation}
for $i=0,1,2$. The same conclusion holds for the
$\mathcal{O}(\epsilon)$ terms but they can be neglected when we take
the $\epsilon\rightarrow 0$ limit. On the other hand, it is well
known that as long as one is concerning about proper ``infrared
safe'' observables, infrared divergences do not show up in the final
result. For example, this can be done by subtracting the IR
divergences from the 1-loop amplitudes via dipole subtracting
scheme~\cite{Catani:1996vz}, and since $C_0,C_1$ and $C_2$ all
vanish in the soft limit, our result is independent of subtraction
schemes. Therefore, we will always refer to our result by stating
its physical implication: for whatever scheme one uses to subtract
the infrared divergences, the one-loop infrared-subtracted
amplitudes for single soft scalar emission always vanish.

\begin{figure}[htp]
 \centering
 \includegraphics[width=0.4\textwidth]{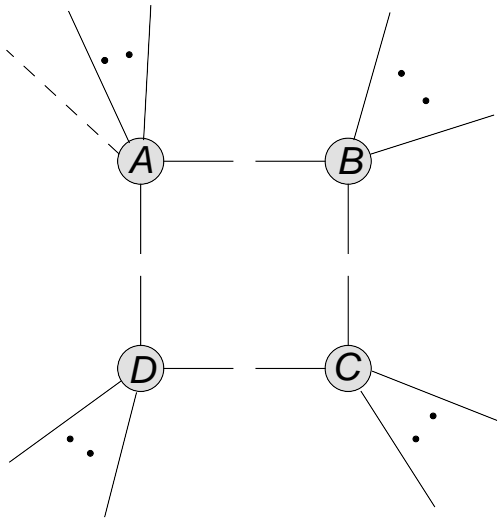}
 \caption{Generic terms of one-loop amplitude with single soft scalar emission.}
 \label{fig:1softgen}
\end{figure}

\begin{figure}[htp]
 \centering
 \includegraphics[width=0.4\textwidth]{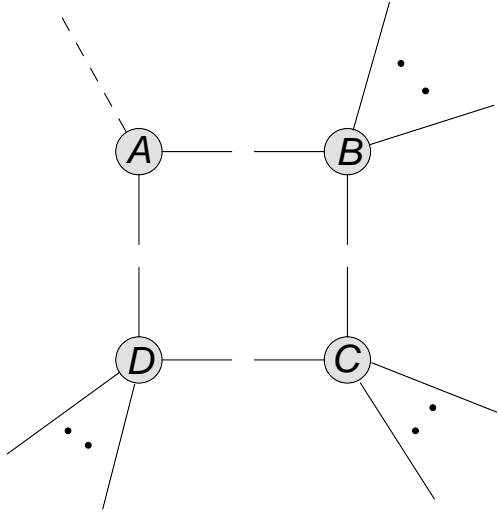}
 \caption{Special terms of one-loop amplitude with single soft scalar emission.}
 \label{fig:1softgen2}
\end{figure}

Now we want to calculate the single soft scalar emission in one-loop
amplitudes, which by Eq.~(\ref{1-loop}) is given by
\begin{eqnarray}\label{n+1}
&&\lim_{p_1\rightarrow 0}\int d^8\eta_{1}
\eta_1^{abcd}\M^{1-loop}_{n+1}=
\nonumber
\\
&&
\lim_{p_1\rightarrow 0}\int
d^8\eta_{1}
\eta_1^{abcd}\sum_{A,B,C,D\subset\{2,...,n+1\}}\sum_{l^*}\prod_{a=AB,BC,CD,DA}\int d^8\eta_{a}\nonumber\\
&&\M_A(\eta_{DA},-l^*_{DA};1;A;\eta_{AB},l^*_{AB})\M_B(\eta_{AB},-l^*_{AB};B;\eta_{BC},l^*_{BC})\nonumber\\
&&\times
\M_C(\eta_{BC},-l^*_{BC};C;\eta_{CD},l^*_{CD})\M_D(\eta_{CD},-l^*_{CD};D;\eta_{DA},l^*_{DA})\nonumber\\
&&\times I_4(P_A+p_1,P_B,P_C,P_D)+\textrm{similar terms}\nonumber\\
&&+\lim_{p_1\rightarrow 0}\int d^8\eta_{1}
\eta_1^{abcd}\sum_{A=\emptyset, B,C,D\subset\{2,...,n+1\}}\sum_{l^*}\prod_{a=AB,BC,CD,DA}\int d^8\eta_{a}\nonumber\\
&&\M_A(\eta_{DA},-l^*_{DA};1;\eta_{AB},l^*_{AB})\M_B(\eta_{AB},-l^*_{AB};B;\eta_{BC},l^*_{BC})\nonumber\\
&&\times
\M_C(\eta_{BC},-l^*_{BC};C;\eta_{CD},l^*_{CD})\M_D(\eta_{CD},-l^*_{CD};D;\eta_{DA},l^*_{DA})\nonumber\\
&&\times I_4(p_1,P_B,P_C,P_D)+\textrm{similar terms},\nonumber\\
\end{eqnarray}
where the first collected term, denoted by $\M_{gen}$
(Fig.~\ref{fig:1softgen}), is the generic case where every corner
has hard external particles and the second one, $\M_{spe}$
(Fig.~\ref{fig:1softgen2}), is the special case where one certain
corner has only a single soft external particle. Besides, similar
terms are those with the soft scalar in corner $B,C$ and $D$. By
Eq.~(\ref{singletree}), we have
\begin{eqnarray}
&&\M_{gen}=\sum_{A,B,C,D\subset\{2,...,n+1\}}\sum_{l^*}\prod_{a=AB,BC,CD,DA}\int d^8\eta_{a}\nonumber\\
&&\left[\lim_{p_1\rightarrow 0}\int d^8\eta_{1}
\eta_{1}^{abcd}\M_A(\eta_{DA},-l^*_{DA};1;A;\eta_{AB},l^*_{AB})\right]
\M_B(\eta_{AB},-l^*_{AB};B;\eta_{BC},l^*_{BC})\nonumber\\
&&\times
\M_C(\eta_{BC},-l^*_{BC};C;\eta_{CD},l^*_{CD})\M_D(\eta_{CD},-l^*_{CD};D;\eta_{DA},l^*_{DA})\nonumber\\
&&\times \lim_{p_1\rightarrow 0}I_4(P_A+p_1,P_B,P_C,P_D)+\textrm{similar terms}.\nonumber\\
\end{eqnarray}
where the soft limit in square parenthesis
vanishes($\mathcal{O}(\delta)$ as $\delta\rightarrow 0$) and other
tree amplitudes are regular since they generally do not depend on
the soft limit(Notice the frozen momenta are generally hard in this
case).

It is easy to see that $\textrm{det}S$ remains regular in the soft
limit, but there are soft momentum divergences arising in the box
functions. Here we only need to consider in $M_{gen}$ box functions
with at least one mass and check their soft limits with a massive
corner $P+p_1$ goes to $P$ as $p_1\rightarrow 0$. If $P$ is massive,
it is easy to see that any of these box functions is regular when
$p_1\rightarrow 0$. For $P$ massless, that is when the corner has
only one soft scalar and a single hard particle, we need to check
potential discontinuities of transitions between $m$-mass functions
and $(m-1)$-mass functions for $m=1,2,3,4$ when taking the soft
limit. As discussed in details in~\cite{Bern:1995ix}, one-mass and
two-mass-easy functions smoothly goes to zero-mass and one-mass
functions in the soft limit, respectively, but two-mass-hard,
three-mass and four-mass functions can have discontinuities
proportional to $1/\epsilon$ in the soft limit. Nevertheless, the
coefficients, which contribute to $C_1$ only diverge as
$\mathcal{O}(\ln(\delta))$ as $p_1\sim
\mathcal{O}(\delta)\rightarrow 0 $, which are overcompensated by the
$\mathcal{O}(\delta)$ vanishing behavior of the product of tree
amplitudes.

Now the only non-trivial thing one needs to check is $\M_{spe}$
\begin{eqnarray}\label{Mspe}
&&\M_{spe}=\sum_{A=\emptyset, B,C,D\subset\{2,...,n+1\}}\sum_{l^*}\prod_{a=AB,BC,CD,DA}\int d^8\eta_{a}\nonumber\\
&&\left[\lim_{p_1\rightarrow 0}\int d^8\eta_{1}
\eta_{1}^{abcd}\M_A(\eta_{DA},-l^*_{DA};1;\eta_{AB},l^*_{AB})\right]
\M_B(\eta_{AB},-l^*_{AB};B;\eta_{BC},l^*_{BC})\nonumber\\
&&\times
\M_C(\eta_{BC},-l^*_{BC};C;\eta_{CD},l^*_{CD})\M_D(\eta_{CD},-l^*_{CD};D;\eta_{DA},l^*_{DA})\nonumber\\
&&\times \lim_{p_1\rightarrow
0}I_4(p_1,P_B,P_C,P_D)+\textrm{similar terms}.\nonumber\\
\end{eqnarray}
It is enough to consider the three-, two-, one-mass cases for $n>3$
and we leave the special case $n=3$ corresponding to the zero-mass
case later for an explicit estimate. For $n>3$, the solution to
Eq.~(\ref{frozen}) is generally hard($\mathcal{O}(1)$), thus we can
use Eq.~(\ref{singletree3})to obtain the soft limit in square
parenthesis, which vanishes as $\mathcal{O}(\delta^2)$, and we now
explicitly estimate the soft limit of box integral when the momentum
of a corner goes to zero.

By Eq.~(\ref{S}), the soft limit of $\textrm{det}S^{-1/2}$ depends
on whether these momenta are on-shell, and it is straightforward to
obtain $\textrm{det}S^{-1/2}\sim \mathcal{O}(1)$ for three-mass and
two-mass-easy cases, while $\textrm{det}S^{-1/2}\sim
\mathcal{O}(\delta^{-1})$ for two-mass-hard, and one-mass cases. For
the box functions defined in Eq.~(\ref{box}), we have the following
results of their soft limits.

For three-mass case, the box function is $\mathcal{O}(1)$ since $s$
goes to a finite value $K_2^2$ when $k_1\rightarrow 0$ and there is
no singular contribution in this soft limit, so does two-mass-easy
case since $s\rightarrow K^2_2$ and $t\rightarrow K^2_4$ in any soft
limit. For two-mass-hard case, there can be singular terms from
$-\frac{1}{\epsilon^2}(-s)^{-\epsilon}$, and
$\frac{1}{2}\ln^2(\frac{s}{t})-\frac{1}{2}\ln\left(\frac{K_4^2}{s}\right)\ln\left(\frac{K_3^2}{s}\right)$
which give $\mathcal{O}(\ln^2\delta)$. Similarly, there can be
singular terms in one-mass case which are
$\mathcal{O}(\ln^2\delta)$.

Therefore, for $n>3$, the soft limit of the box integrals
$\lim_{p_1\rightarrow 0}I_4(p_1,P_B,P_C,P_D)$ can have soft momentum
divergences, but these are overcompensated by the three particle
amplitude which vanishes as $\mathcal{O}(\delta^2)$ in the soft
limit, and we obtain that
\begin{eqnarray}\label{n>3}
\lim_{p_1\rightarrow 0}\int d^8\eta_{1}
\eta_{1}^{abcd}\M^{1-loop}_{n+1}(1,2,...,n+1)=0
\end{eqnarray} for
$n>3$.

\begin{figure}[htp]
 \centering
 \includegraphics[width=0.4\textwidth]{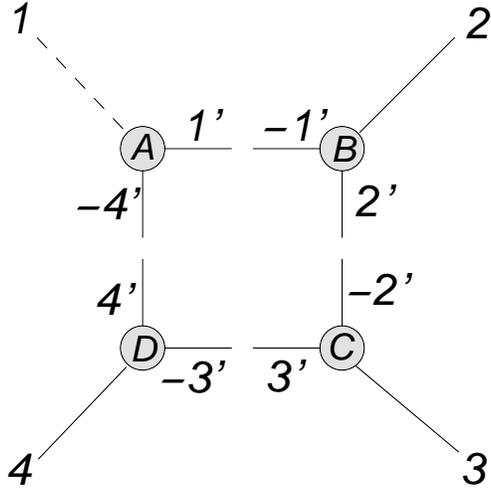}
 \caption{One-loop 4-point amplitudes with one soft particle and three hard particle.}
 \label{fig:sMMM}
\end{figure}

The case $n=3$ is more subtle and we treat it separately
here(Fig.~\ref{fig:sMMM}). Take the term with $p_1=P_A$ in the
corner A as an example and relabel
$P_A=k_1\sim\delta$,$P_B=k_2$,$P_C=k_3$ and $P_D=k_4$. Since
$k_1,k_2,k_3$ and $k_4$ are all on-shell, in the soft limit
$k_1\sim\delta\rightarrow 0$, we have not only $s=2k_1\cdot
k_2\sim\delta$, but also $t=2k_1\cdot k_4\sim\delta$. Therefore, the
pre-factor $\textrm{det}S^{-1/2}\sim \mathcal{O}(\delta^{-2})$ and
the corresponding zero-mass box function is
$\mathcal{O}(\ln^2\delta)$ in this limit. In addition, we can not
just take the soft limit inside square parenthesis of
Eq.~(\ref{Mspe}) because in this case some internal(fixed) momenta
$l^*$ become soft!

To see this, we shall use Eq.~(\ref{frozen}) which gives,
\begin{equation}
l^*\cdot k_1=l^*\cdot k_2=l^*k_4-k_2\cdot k_3=0,
\end{equation}
where in the last equality $l^*\cdot k_2=k_2\cdot k_3=-k_1\cdot
k_4\sim \mathcal{O}(\delta)$ implies that $l^*\sim
\mathcal{O}(\delta)$ and further $(l^*+k_1)\sim
\mathcal{O}(\delta)$, thus three momenta $-l^*_{DA},p_4$ and
$l^*_{AB}$ in $\M_A$, $-l^*_{AB}$ in $\M_B$ and $l^*_{DA}$in $\M_D$
are all $\mathcal{O}(\delta)$. Adopting a simpler notation, we have
\begin{eqnarray}
\label{M4}
&&\int d^8\eta_1\eta_1^{abcd}\M_4(1,2,3,4)=\nonumber\\
&&\int d^8\eta_1\eta_1^{abcd}\prod^{4}_{a=1}\int d^8\eta_{a'}
\M_A(-4',1,1')\M_B(-1',2,2')\M_C(-2',3,3')\M_D(-3',4,4')\nonumber\\
&&\times I_4(1,2,3,4)+\textrm{similar terms}.\nonumber\\
\end{eqnarray}

First we assume that the sub-amplitude $\M_A$ is anti-holomorphic.
The similar argument can be applied to making the opposite choice of
$\M_A$ be holomorphic. The sub-amplitudes $\M_B$, $\M_C$ and $\M_D$
can be either holomorphic or anti-holomorphic. But it will be clear
later that the case where all of the sub-amplitudes are
anti-holomorphic is irrelevant for single soft scalar emission. For
all the sub-amplitudes, we choose to work in $\eta$ representation.
In this representation three particle amplitude is
\begin{equation}
 \M^{ah}_3(\eta_i)=\frac{\delta^{16}(\tilde{\lambda}_i\eta_i)}
{(\spb 1.2 \spb 2.3 \spb 3.4)^2}
\end{equation}
for $\M_3$ anti-holomorphic and
\begin{equation}
\M^h_3(\eta_i)=\int~\prod_{i=1,2,3} d^8\bar{\eta}_{i}
e^{\bar{\eta}_i\eta_i}
\frac{\delta^{16}(\lambda_i\bar{\eta}_i)}{(\spa 1.2 \spa 2.3 \spa
3.1)^2}
\end{equation}
for $\M_3$ holomorphic. Note that in $\eta$ representation, the
power of $\eta$ is 16 for an anti-holomorphic amplitude and 8 for a
holomorphic amplitude. We define an useful quantity
$\varDelta(\M_n(\eta))$ to be the power of $\eta$ minus the power of
$d\eta$ in $\M_n$. Obviously, $0 \leq \varDelta(\M_n)\leq 8n$ For
any amplitude to be non-vanishing.
It is easy to see that for the amplitude defined in Eq.~(\ref{M4}),
we have
\begin{equation}
 \varDelta(\M_4)=\left\{
\begin{array}{l l}
 32, & \textrm{if the number of holomorphic amplitude in } \{\M_B,\M_C,\M_D\}
\textrm{ is } 0,
\\
24, & \textrm{if the number of holomorphic amplitude in }
\{\M_B,\M_C,\M_D\} \textrm{ is } 1,
\\
16, & \textrm{if the number of holomorphic amplitude in }
\{\M_B,\M_C,\M_D\} \textrm{ is } 2,
\\
8, & \textrm{if the number of holomorphic amplitude in }
\{\M_B,\M_C,\M_D\} \textrm{ is } 3.

\end{array}
\right.
\end{equation}
Now we can understand why the case when all of the sub-amplitudes
are anti-holomorphic is irrelevant. In this case $\varDelta(\M_4)
=32$, when taking into account the $4\times8$ fold $d\eta$ integral,
the only choice for the external particles species is 4 gravitons!
If two of the sub-amplitudes are anti-holomorphic, $\varDelta=16$
and there are $4\times8$ fold $d\eta$ external particle integral,
left 8 power of $\eta$ assignment for external particle species. For
the bosonic case, it can correspond to cases studied in~\cite{KLR},
a 4-point amplitude of two scalars and two vectors, or that of four
scalars.

Since $\M_A$ is anti-holomorphic, we can set $\lambda_{4'}$,
$\lambda_1$ and $\lambda_{1'}$ to be parallel and take the soft
limit as
$$\lambda_{4'}=\lambda_{1'}=\lambda_1=\delta^\alpha\hat{\lambda}_1$$,
and
$$\tlambda_{4'}=\delta^{1-\alpha}\hat{\tlambda}_{4'},\qquad
\tlambda_{1'}=\delta^{1-\alpha}\hat{\tlambda}_{1'},\qquad
\tlambda_1=\delta^{1-\alpha}\hat{\tlambda}_1.
$$
By momentum conservation we have
$\tilde{\lambda}_1=\tilde{\lambda}_{4'}-\tilde{\lambda}_{1'}$. Then
$\M_A$ is given by
\begin{equation}
 \M_A=\delta^{4-4\alpha}{\spb {\hat{4}'}.{{\hat{1}'}}}^2
\delta^8(\eta_{1'}-\eta_1)\delta^8(\eta_{4'}+\eta_1)\sim
\mathcal{O}(\delta^{4-4\alpha}).
\end{equation}
The power of $\eta_1$ in $\M_A$ is 4 in order to match the
pre-factor $\eta^{abcd}$ when considering a scalar emission. Thus in
total $\M_A$ provides 12 power of $\eta_{4'}$ plus $\eta_{1'}$. In
order to match the internal $d\eta_{4'}$ and $d\eta_{1'}$ integral,
$\M_B$ and $\M_D$ must provide 4 power of $\eta_{4'}$ plus
$\eta_{1'}$.

The holomorphic part of $\M_B$ is given by
\begin{equation}
 \M^{h}_B=\prod_i\int\,d^8\bar{\eta}_i e^{\bar{\eta}_i\eta_i}
\delta^{-2+4\alpha}{\spa {\hat{1}'}.{2'}}^2
\delta^8(\bar{\eta}_{1'}+\delta^{1-\alpha}\bar{\eta}_2)
\delta^8(\bar{\eta}_{2'}-\bar{\eta}_2),
\end{equation}
and the anti-holomorphic part is
\begin{equation}
 \M^{ah}_B=\delta^{2-4\alpha}{\spb {\hat{1}'}.{2'}}^2
\delta^8(\eta_{1'}+\delta^\alpha\eta_2)\delta^8(\eta_{2'}+\eta_2).
\end{equation}
Besides, $\M^{h(ah)}_D=\M^{h(ah)}_B(1'\rightarrow 3',2'\rightarrow
4', 2\rightarrow 4)$. There is no soft momentum in the other
holomorphic sub-amplitude $\M_D$, thus it is always
$\mathcal{O}(1)$.

We discuss several choices for $\M_B$ and $\M_D$. If both $\M_B$ and
$\M_D$ are holomorphic, we have
\begin{eqnarray}
 \M_4&\sim& \int\,d^8\eta_1 \eta^{abcd} d^8\eta_{1'} d^8\eta_{4'}
\delta^{4-4\alpha}{\spb {\hat{4}'}.{{\hat{1}'}}}^2
\delta^8(\eta_1'-\eta_1)\delta^8(\eta_4'+\eta_1) \nonumber
\\
&&\times \prod_i\int\,d^8\bar{\eta}_i e^{\bar{\eta}_i\eta_i}
\delta^{-2+4\alpha}{\spa {\hat{1}'}.{2'}}^2
\delta^8(\bar{\eta}_{1'}+\delta^{1-\alpha}\bar{\eta}_2)
\delta^8(\bar{\eta}_{2'}-\bar{\eta}_2) \nonumber
\\
&&\times \prod_i\int\,d^8\bar{\eta}_i e^{\bar{\eta}_i\eta_i}
\delta^{-2+4\alpha}{\spa {\hat{3}'}.{4'}}^2
\delta^8(\bar{\eta}_{4'}-\delta^{1-\alpha}\bar{\eta}_4)
\delta^8(\bar{\eta}_{3'}+\bar{\eta}_4) \nonumber
\\
\label{MBhMDh} &&\times\delta^{-2}\log^2\delta.
\end{eqnarray}

As mentioned before, $\M_B$ and $\M_D$ must provide 4 power of
$\eta_{1'}$ plus $\eta_{4'}$. This can only come from the
exponential of second and third line in Eq.~(\ref{MBhMDh}). At the
same time it brings down 4 power of $\bar{\eta}_{1'}$ plus
$\bar{\eta}_{4'}$ from the exponential. In order to match the
$d\bar{\eta}_{1'}$ and $d\bar{\eta}_{4'}$ integral, there must be 12
power of $\bar{\eta}_{1'}$ plus $\bar{\eta}_{4'}$ from the delta
functions in the second and third line in Eq.~(\ref{MBhMDh}). Thus
the same delta functions provide $\delta^{4-4\alpha}$,
\begin{eqnarray}
 \M_4&\sim&\delta^{4-4\alpha}\cdot\delta^{-2+4\alpha}\cdot
\delta^{-2+4\alpha}\cdot\delta^{4-4\alpha}\cdot
\delta^{-2}\log^2\delta\nonumber
\\
&\sim&\delta^2\log^2\delta.
\end{eqnarray}
This establishes that $\M_4$ vanishes as $\delta\rightarrow 0$.

Next we consider the case where $\M_D$ is holomorphic and $\M_B$ is
anti-holomorphic. The opposite choice is similar.
\begin{eqnarray}
\label{aha}
  \M_4&\sim& \int\,d^8\eta_1 \eta^{abcd} d^8\eta_{1'} d^8\eta_{4'}
\delta^{4-4\alpha}{\spb {\hat{4}'}.{{\hat{1}'}}}^2
\delta^8(\eta_1'-\eta_1)\delta^8(\eta_4'+\eta_1) \nonumber
\\
&&\times\delta^{2-4\alpha}{\spb {\hat{1}'}.{2'}}^2
\delta^8(\eta_{1'}+\delta^\alpha\eta_2)\delta^8(\eta_{2'}-\eta_2)
\\
\nonumber &&\times \prod_i\int\,d^8\bar{\eta}_i
e^{\bar{\eta}_i\eta_i} \delta^{-2+4\alpha}{\spa {\hat{3}'}.{4'}}^2
\delta^8(\bar{\eta}_{4'}-\delta^{1-\alpha}\bar{\eta}_4)
\delta^8(\bar{\eta}_{3'}+\bar{\eta}_4) \nonumber
\\
&&\times\delta^{-2}\log^2\delta.
\end{eqnarray}

The $\delta$ power counting reads
\begin{eqnarray}
\M_4&\sim&
\delta^{4-4\alpha}\cdot\delta^{2-4\alpha}\cdot\delta^{-2+4\alpha}
\cdot\delta^{(8-j)\alpha}\cdot\delta^{(4-j)(1-\alpha)}\cdot\delta^{-2}
\cdot\log^2\delta \nonumber
\\
&\sim&\delta^{6-j}\log^2\delta,
\end{eqnarray}
where $j$ is the number of $\eta_{1'}$ in Eq.~(\ref{aha}). The
dominant contribution comes form taking $j=4$, and $\M_4$ scales as
$\delta^2\log\delta$.

If both $\M_B$ and $\M_D$ are anti-holomorphic, we have
\begin{eqnarray}
  \M_4&\sim& \int\,d^8\eta_1 \eta^{abcd} d^8\eta_{1'} d^8\eta_{4'}
\delta^{4-4\alpha}{\spb {\hat{4}'}.{{\hat{1}'}}}^2
\delta^8(\eta_1'-\eta_1)\delta^8(\eta_4'+\eta_1) \nonumber
\\
&&\times\delta^{2-4\alpha}{\spb {\hat{1}'}.{2'}}^2
\delta^8(\eta_{1'}+\delta^\alpha\eta_2)\delta^8(\eta_{2'}-\eta_2)
\nonumber
\\
&&\times\delta^{2-4\alpha}{\spb {\hat{4}'}.{3'}}^2
\delta^8(\eta_{4'}-\delta^\alpha\eta_4)\delta^8(\eta_{3'}+\eta_4)
\nonumber
\\
&&\times\delta^{-2}\log^2\delta \nonumber
\\
&\sim&\delta^6\log^2\delta\rightarrow 0.
\end{eqnarray}

It is clear that our result is in agreement with the result derived
in~\cite{KLR} for $\varDelta(\M_4)=16$, although it is more general
because it is applicable directly to cases with fermions. In
addition, our result holds for cases with $\varDelta(\M_4)=8,24$
because we did not assume wether $\M_C$ is holomorphic or
anti-holomorphic(except the case when both $\M_B$ and $\M_D$ are
anti-holomorphic for which it must be holomorphic), thus our result
covers all possible arrangement of four external particles with at
least one scalar and it shows that one-loop four-point amplitudes
with single soft scalar emission vanish in all cases. Together with
Eq.~(\ref{n>3}), the conclusion is, for the infrared-subtracted
amplitude,
\begin{eqnarray}\label{singleloop}
\lim_{p_1\rightarrow 0}\int d^8\eta_{1}
\eta_{1}^{abcd}\M^{1-loop}_{n+1}(1,2,...,n+1)=0
\end{eqnarray}
for $n\geq3$.

Naively we should be able to generalize our result to one-loop
double soft scalar emission. At the tree level, authors
of~\cite{ACK} have obtained a finite result which reveals the
non-trivial structure of $E_{7(7)}$ group. It is expected that the
same result at the one-loop level should directly follows from
Eq.(\ref{1-loop}) and the vanishing result of the tree level single
soft emission. However, the discontinuities between different box
functions make the problem non-trivial since we must explicitly take
into account discontinuities to check if the same finite result can
be obtained at the one-loop level. This work is in
progress~\cite{HZ2}.

\section{Conclusion and Discussions}
In this note we have studied single soft scalar emission of
$\Neqeight$ SUGRA. As investigated from different perspectives
in\cite{BEF,ACK,KK}, at the tree level, the single soft scalar
emission vanishes which indicates a hidden $E_{7(7)}$ symmetry in
addition to the $SU(8)$ symmetry. Here we generalize the result to
the one-loop level using supersymmetric BCFW construction and a
simple formula for one-loop amplitude of $\Neqeight$ SUGRA in terms
of tree amplitudes, due to the absence of triangle, bubble and
rational terms. It turns out that for the one-loop
infrared-subtracted amplitude, the single soft scalar emission
vanishes, which implies that there may be no anomalies of the
$E_{7(7)}$ symmetry at the one loop level.

Our result is in agreement with that of~\cite{KLR} for special cases
studied there, i.e. four-scalar and two-scalar-two-vector
amplitudes. Although we have not obtained the explicit expression
for general amplitudes as for special cases in~\cite{KLR}, the
vanishing result for infrared finite parts of general amplitudes is
obtained for the first time. As argued in~\cite{KLR}, this should
directly imply the ``axial'' charge vanishes,
which by the low energy theorem implies the
conservation of the corresponding Noether current.

Clearly more works are needed to reveal the role of $E_{7(7)}$ and
possible enlarged symmetry in $\Neqeight$ SUGRA. First, by analyzing
the subtraction of infrared divergences properly, it is
straightforward to study the double soft scalar emission at the
one-loop level to further confirm the non-trivial structure of
$E_{7(7)}$ group obtained by double emission at the tree
level~\cite{ACK}. Furthermore, it would be very interesting to study
the soft emission of arbitrary numbers of scalars at both the tree
and loop level, and the results should indicate the exponentiation
and the full finite action of $E_{7(7)}$ group on the Hilbert space.
Besides, as discussed in ~\cite{ACK}, the single soft emission of
graviphoton can go to a constant, which may indicate further
enlarged symmetry of the theory, and further investigations for such
emissions at both tree and loop level are desirable. We hope that
the result of soft scalar and graviphoton emissions, which reveals
$E_{7(7)}$ and possible enlarged symmetry, can shed some light on
the possible UV finiteness of $\Neqeight$ SUGRA.

\section*{Acknowledgement}
We are grateful to J.~Kaplan for helpful discussions. S.H. thanks
N.~Arkani-Hamed and F.~Cachazo for encouragement on working along
this direction. H.Z. is grateful to Chong Sheng Li for his support
on this work. S.H.'s work is supported by the National Natural
Science Foundation (NFS) of China under grant No. 10721063, No.
10675005 and No. 10835002. H.Z. is supported by National Natural
Science Foundation of China, under Grants No.10721063, No.10575001
and No.10635030.

\end{document}